\documentclass{article}

\usepackage{graphicx}
\usepackage{psfig}
\usepackage{epsfig}
\usepackage[round]{natbib}

\title{Moment based methods for ensemble assessment and calibration}

\author{Stephen Jewson\footnote{\emph{Correspondence address}: RMS, 10 Eastcheap,
London, EC3M 1AJ, UK. Email: \texttt{x@stephenjewson.com}}\\
RMS, London, United Kingdom}

\begin{document}

\newcommand{\bx}[1]{\fbox{\begin{minipage}{15.8cm}#1\end{minipage}}}

\maketitle

\begin{abstract}

We describe various moment-based ensemble interpretation models for the construction 
of probabilistic temperature forecasts from ensembles.
We apply the methods to one year of medium range ensemble forecasts
and perform in and out of sample testing. Our main conclusion is that 
probabilistic forecasts derived from the ensemble mean using regression are just as good
as those based on the ensemble mean and the ensemble spread using a more complex calibration algorithm. 
The explanation for this
seems to be that the predictable component of the variability of the forecast uncertainty is only a small fraction
of the total forecast uncertainty.
Users of ensemble temperature forecasts are advised, until further evidence becomes available, 
to ignore the ensemble spread and build probabilistic
forecasts based on the ensemble mean alone.

\end{abstract}

\section{Introduction}

We consider the question of how to make probabilistic forecasts of site-specific temperatures.
An interesting subset of this problem considers the probabilities of temperature falling into
a discrete number of bins or categories. However, we are interested in the more general
question of how to make a prediction of the continuous distribution of possible temperatures.
The continuous distribution case is the case most appropriate for users 
who make frequent use of forecasts in continuously changing situations, and who cannot define any
specific thresholds or categories of temperature as being important in their application.
Examples include most users of forecasts in the energy and weather derivatives industries.  
This is not to say that users of forecasts in these industries do not
also, on occasion, use categorical forecasts, but rather that most commonly they simply want
the best possible probabilistic forecast, without further specialisation.

\citet{jewsonbz03a} (henceforth JBZ)
propose the use of moment-based methods for the assessment and calibration of ensemble forecasts 
within the continuous distribution framework. This paper discusses these methods in more detail,
and compares a number of different moment-based algorithms.  
We then test the methods using both in and out of sample testing, and draw clear conclusions about their
relative usefulness. We follow the guidelines described by~\citet{jewsonz03b}, which specify
a set of necessary conditions for a methodology to satisfy if one wants to draw conclusions about the
usefulness of real forecasts in real applications. 
Finally, we come to some surprising conclusions about the value of
the ensemble spread in making probabilistic forecasts. 

This paper is part of a wider research program in which the authors are attempting to understand the
potential uses for meteorological forecasts in the weather derivative industry.
Within this context, moment-based ensemble calibration methods were introduced for the following reasons:

\begin{itemize}
  \item To provide a simple answer to the question of how many days of forecast are better than climatology,
  for both the ensemble mean and the standard deviation, and to give clear statistical tests to help answer
  this question.
  \item To create a method for calibration of the ensemble that is both simpler and more accurate
  than the rather cumbersome methods currently in use such as the rank histogram.
  In particular, to derive a method that optimally merges estimates of the forecast uncertainty 
  based on past forecast error
  statistics with those based on the ensemble spread.
\end{itemize}

\section{Data}
\label{data}

We will base our analyses on one year of ensemble forecast data for the weather
station at London's Heathrow airport, WMO number 03772. The forecasts are predictions
of the daily average temperature, and the target days of the forecasts
run from 1st January 2002 to 31st December 2002. The forecast was produced
from the ECMWF model~\citep{molteniet96} and downscaled to the airport location using a simple
interpolation routine prior to our analysis. There are 51 members in the ensemble.
We will compare these forecasts to the quality controlled climate
values of daily average temperature for the same location as reported by the UKMO.

There is no guarantee that the forecast system was held constant throughout this period,
and as a result there is no guarantee that the forecasts are in any sense stationary,
quite apart from issues of seasonality. This is clearly far from ideal with respect to 
our attempts to build statistical interpretation models on past forecast data but is,
however, unavoidable: this is the data we have to work with.

Throughout this paper all equations and all values have had both the seasonal mean and
the seasonal standard deviation removed. Removing the seasonal standard deviation
removes most of the seasonality in the forecast error statistics, and partly justifies the use of
non-seasonal parameters in the statistical models for temperature that we propose.

\section{Moment-based ensemble calibration methods}
\label{model}

Perhaps the simplest way to make a reasonable probabilistic forecast of temperature is to build a 
least-squares minimising linear regression model between past forecasts and observations. 
Typically in such a model one might
assume that the forecast errors follow a normal distribution and are uncorrelated in time. 
The model for temperature $T_i$ on day $i$ can then be written as:

\begin{equation}\label{regression}
  T_i \sim N(\hat{\alpha}+\hat{\beta} m_i,\hat{\gamma})
\end{equation}

where $m_i$ is the forecast and $\hat{\alpha}, \hat{\beta}, \hat{\gamma}$ are constants to be determined
as part of the calibration. The reason we have used the symbol $\gamma$ rather than the more usual $\sigma$ will
become apparent below.
In the ensemble forecast case $m_i$ would be the ensemble mean.
Least squares linear regression provides analytical 
expressions for the parameters $\hat{\alpha}, \hat{\beta}, \hat{\gamma}$ in terms of the data (see~\citet{NR}). 
The regression-based ensemble calibration model says that the ensemble mean $m_i$ is converted 
into an estimate for the mean of the forecast distribution $\hat{\mu}$ using the
linear transformation $\hat{\alpha}+\hat{\beta} m_i$ which both corrects a bias in the forecast and also damps the
forecast towards climatology in such a way as to minimise root mean square error. 

The use of least squares minimisation for the fitting of the parameters in the regression model is a special
case of the use of the more general maximum likelihood method of~\citet{fisher1922}.
Maximum likelihood is the most common
parameter fitting method used in statistics (see, for example, \cite{casellab02}).
It works as follows:
\begin{itemize}
  \item The likelihood is defined as the probability density of the entire observed data set given the
  forecasts and a particular set of
  parameters for the model. This probability is a single number.
  \item The parameters are then varied. This changes the likelihood.
  \item The optimum parameter values are chosen to be those that maximise the likelihood, i.e. maximise the
  probability density of the data given the parameters and the forecast.
\end{itemize}

The likelihood is a natural way to define the goodness of fit between a model and a data set.~\footnote
{We note that our use of the word \emph{likelihood} follows the original definition 
from classical statistics.
Occassionally, and rather confusingly, the word has been used in a slightly different sense in meteorology.
We discuss this question in~\cite{jewson03f}}
However, it is also much
more: it can be shown that, for a very large class of models that includes most reasonable models (and certainly
all the models we will consider below) use of the maximum likelihood method
gives the most accurate possible estimates for the parameters.

In the case of equation~\ref{regression} the likelihood can be derived from the probability density function for the 
multivariate normal distribution, and is given by:

\begin{equation}\label{lik}
  L=\frac{1}{\sqrt{2 \pi \mbox{det}}}\mbox{exp}(-\frac{1}{2}(T-\hat{\mu})^T \hat{\Sigma}^{-1} (T-\hat{\mu}))
  \end{equation}

where $\hat{\Sigma}$ is the estimated covariance matrix of the forecast errors, $T$ is the vector of observations, 
$\hat{\mu}=\hat{\alpha}+\hat{\beta} m$ is the vector of forecast temperatures and $\mbox{det}$ is the determinant
of $\hat{\Sigma}$.
We see that $L$ is a function of both the observed data (via $T$),
the forecasts (via $\hat{\mu}$, $\hat{\Sigma}$ and det) and the parameters of the model 
(also via $\hat{\mu}$, $\Sigma$ and det).
In practice it is often more convenient to consider the log-likelihood rather than the likelihood.  
Since the log function is monotonically increasing, finding the parameters that maximise one
is equivalent to finding the parameters that maximise the other.
The log-likelihood in this case is given by:

\begin{equation}\label{loglik1}
  l=-\frac{1}{2} ln (2 \pi \mbox{det}) - \frac{1}{2}(T-\hat{\mu})^T \hat{\Sigma}^{-1} (T-\hat{\mu})
\end{equation}

If we make the assumption that the forecast errors are uncorrelated in time then $\hat{\Sigma}$ is
diagonal and this simplifies to:

\begin{equation}\label{loglik2}
  l = -\frac{1}{2} \sum_{i=1}^{i=n} ln (2 \pi \hat{\sigma}) 
    - \frac{1}{2} \sum_{i=1}^{i=n} \frac{(T_i-\hat{\mu_i})^2}{\hat{\sigma}^2}
\end{equation}

The values of $\hat{\alpha}, \hat{\beta}, \hat{\sigma}$ that maximise this expression can be derived analytically, and
yield the expressions that are used to give the parameters in least-squares linear regression. 
Expressions for the sampling uncertainty
can also be derived analytically since it can be shown that the approximate sampling uncertainty on the parameters
is given by the inverse of the matrix of second derivatives of the log-likelihood.

The biggest shortcoming of the regression model as a method for creating probabilistic forecasts is that it
cannot predict flow-dependent changes in the uncertainty, since the uncertainty $\hat{\gamma}$ is a constant. 
It is well established
that the ensemble spread contains information about this uncertainty
and it would seem sensible to extend the
regression model to attempt to incorporate some of this information. We will present three such extensions. 
The first simply uses the ensemble spread $s_i$ to predict the uncertainty, rather than deriving the uncertainty
estimate from past forecasts. This is now a two-parameter model, which we write as:

\begin{equation}\label{spreadonly}
  T_i \sim N(\hat{\alpha}+\hat{\beta} m_i, s_i)
\end{equation}

We will call this the \emph{spread only} model.

The third model assumes that the ensemble spread is a good predictor of the uncertainty, but that it
may have the wrong amplitude, and so we add a parameter to scale it. This is a three-parameter model 
and is written as:
\begin{equation}\label{spreadscaling}
  T_i \sim N(\hat{\alpha}+\hat{\beta} m_i, \hat{\delta} s_i)
\end{equation}

We will call this the \emph{spread-scaling} model.
This model suffers from what we call the "spread-inflation problem", which is that
the mean level of the spread and the variability of the spread are scaled by the same factor.
This is clearly not the best way to optimise forecasts. To take a limiting case as illustration:
if the variability in the spread is meaningless then we should set it to zero, but we should
certainly not set
the mean spread to zero. The spread-inflation problem is also present in the 
rank histogram and best-member~\citep{roulstons03} ensemble
calibration methods, which is why we do not consider them in our analysis.

Our fourth model is the \emph{spread-regression} model from JBZ, 
which overcomes the spread-inflation problem at the cost of introducing a fourth parameter:

\begin{equation}\label{jbz}
  T_i \sim N(\hat{\alpha}+\hat{\beta} m_i, \hat{\gamma} + \hat{\delta} s_i)
\end{equation}

In this model, if the ensemble spread contains no information then $\hat{\delta}$ can be zero and the uncertainty is
modelled entirely on the basis of past forecast error statistics as in equation~\ref{regression}. 
This idea that the spread might contain no information is not just a thought experiment:
~\citet{jewsondh03a} give some examples of ensembles in which it appears that there is no skill in the spread, and for
which $\hat{\delta}$ is thus zero.

The $\hat{\gamma}$ and $\hat{\delta}$ parameters in the spread-regression model can be interpreted as follows. 
If $\hat{\delta}$ is not significantly different from
one then the variability in the ensemble spread has the correct amplitude and does not need scaling. 
If $\hat{\delta}$ is not
significantly different from zero then the variability in the spread has no information in the context of this
interpretation model, and should be ignored. This was the case for long leads in the forecasts discussed by 
\citet{jewsondh03a}. 

Forecasts derived from general circulation models contain no useful information \emph{at all} about either the
average site-specific temperature or the average level of uncertainty of forecasts of that temperature, 
since both quantities are 
highly affected by local conditions not represented in the model. The purpose of the coefficients
$\hat{\alpha}$ and $\hat{\gamma}$ is to use site-specific observational data to set these averages levels of temperature
and uncertainty correctly. What general circulation models do potentially represent well are the fluctuations in 
temperature and the fluctuations in uncertainty of the forecast: these are captured by the $\hat{\beta}$ and $\hat{\delta}$ terms.

There are various other values that can be derived from the parameters in equation~\ref{jbz} that give further insight
into the performance of the ensemble. One is the ratio of the mean spread after calibration to the mean spread
before calibration. This is given by:

\begin{equation}
 \mbox{MSR}=\mbox{mean spread ratio}= \frac{\hat{\gamma}+\hat{\delta} \overline{s}}{\overline{s}}
\end{equation}

and tells us to what extent the ensemble underestimates the average level of the spread.
Another useful diagnostic is the coefficient of variance of spread before and after calibration, given by:

\begin{equation}
  \mbox{COVS}_{\mbox{before}}=\frac{\mbox{sd}(s)}{\overline{s}}
\end{equation}

for the uncalibrated ensemble and 

\begin{equation}
  \mbox{COVS}_{\mbox{after}}=\frac{\hat{\delta} \mbox{sd}(s)}{\hat{\gamma}+\hat{\delta} \overline{s}}
\end{equation}
for the calibrated ensemble.
The COVS tells us the ratio of the size of the variations in the spread to the mean spread.

The third is the \emph{deterministic-stochastic}, or \emph{ds}, ratio. This is defined as:
\begin{equation}
  \mbox{DS ratio}=\frac{\overline{s}}{sd(m)}
\end{equation}
This ratio varies from zero for a perfectly deterministic forecast with no uncertainty, to infinity for
a perfectly stochastic forecast with no variability in the prediction of the mean. This ratio thus tells
us to what extent the forecasts are closer to the deterministic limit or the stochastic limit.

\subsection{Skill scores}

In the context of the regression model the most useful skill score is the root-mean-square-error (RMSE), 
which measures the typical sizes
of errors. If the forecast errors really are normally distributed, and the forecasts have been calibrated
using equation~\ref{regression}, 
then the RMSE tells us everything we could ever need to know about the skill of the forecast.

However, the main difference between the four models discussed above is in their ability to predict 
the uncertainty,
and the RMSE does not take that into account. Rather, we need a measure that considers the ability to predict
the whole distribution of possible outcomes.
JBZ propose the likelihood,
or measures derived from the likelihood, as an appropriate way to make this comparison. 
Advantages of the likelihood are that it avoids having to make arbitrary categories, it incorporates all 
of the observed data and it evaluates over the whole range of the forecast. The advantages of using
the likelihood as a skill measure are discussed in more detail in~\citet{jewson03e}.

There are a number of ways the likelihood can be presented such as
actual likelihood values, the
log-likelihood, the root of minus the mean of the log-likelihood (RMMLL) or as the likelihood skill score.
We will use the RMMLL.

Since the spread-regression model has four
parameters, it is almost guaranteed to perform better than the other models in in-sample data fitting tests. 
But this does not mean it is a better model. One way around this problem is to adjust the in-sample 
likelihood to penalise
each model for the number of parameters used with a factor of $\frac{1}{2}Qln(N)$ where $Q$ is the number
of parameters and $N$ is the number of data points (this is known as the \emph{Bayesian information criterion}, 
or BIC).
We will use the BIC adjustment in the RMMLL scores presented in section~\ref{testing}.
Alternatively we can use out of sample testing. 
We will use a combination of in-sample and out of sample testing in our model comparisons below.

\subsection{Fitting the parameters}

We fit our four models under the assumption that the forecast errors are uncorrelated in time, which
means we can calculate the log-likelihood using equation~\ref{loglik2}. 
We will see later, however, 
that this assumption is not exactly correct, and hence
in principle one should use the log-likelihood given by equation~\ref{loglik1} to fit all the models. 
However, in practice using equation~\ref{loglik1} is somewhat awkward since the
covariance matrix of the forecast errors $\Sigma$ is not known in advance. 
Developing algorithms to fit the four models using equation~\ref{loglik1} is an area for further work.

The parameters in the regression, spread-only and spread-scaling models  
(equations~\ref{regression}, \ref{spreadonly} and \ref{spreadscaling})
can be fitted using analytic expressions that can be derived by differentiating equation~\ref{loglik2}
with respect to the parameters, setting the derivatives equal to zero, and solving with respect
to the parameters. 
This does not appear to be possible
for equation~\ref{jbz}, however, because the resulting equations are not analytically tractable.
Instead, numerical optimisation routines must be used. The simplest effective optimisation
routines are the so-called quasi-Newton schemes (see~\cite{NR}). 
These are simple to use because they only
involve the user having to make calculations of the log-likelihood, and not its derivatives. 
Although deriving analytic expressions for the gradient and curvature is easy for 
equation~\ref{loglik2}, in general it can be very difficult (for example for equation~\ref{loglik1}).

We have found that quasi-Newton schemes are perfectly adequate for calculation of the parameters
in equation~\ref{jbz} and converge within a small number of steps
to the maximum value. This takes only a fraction of a second on even the slowest personal computer.

\subsection{Extensions of the spread-regression model}
\label{extensions}

One could criticize our four models for assuming 
that the forecast errors are normally distributed.
In fact, they can easily be extended to cope with other distributions. 
There are two ways in which this could be done.
First, one could apply linear transformations to the parameters of a different distribution. 
This might be appropriate
if trying to calibrate a gamma distribution fitted to precipitation, for instance. Secondly, one could still apply
linear transforms to the mean and the standard deviation, but then fit a distribution other than the normal,
such as a kernel density. In that case, the model would have a fifth parameter for the bandwidth
of the kernel.
Such a model could again be fitted using maximum likelihood, and may have
the benefit that it would allow a more flexible range of distribution shapes than the normal distribution. Testing
whether such a model really would give benefits over the normal distribution is an area of current research.

\section{Model testing and validation}
\label{testing}

We will now investigate some of the statistical characteristics of our forecast data,
and compare the performance of the four models described above.

\subsection{Characteristics of the ensemble forecasts}

We first look briefly at some of the characteristics of the variability of the 
mean and the standard deviation of the ECMWF ensemble forecast.

\begin{figure}[!htb]
  \begin{center}
    \scalebox{0.9}{\includegraphics{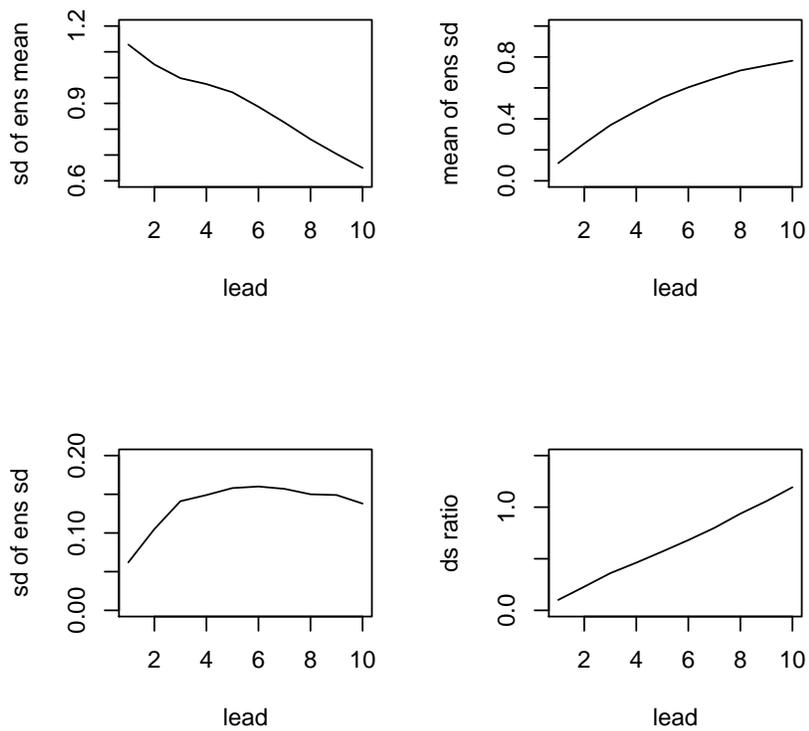}}
  \end{center}
 \caption{
 Diagnostics of the ensemble forecast data used in this study, all for double anomalies,
 and all versus lead.
 Panel a)shows the standard deviation of the ensemble mean, 
 b)shows the mean of the ensemble standard deviation,
 c)shows the standard deviation of the ensemble standard deviation and 
 d)shows the ratio of b) to a), which is a measure of the extent to which
 the ensemble is close to being like a pure deterministic system (low values)
 or a pure stochastic system (high values).}
 \label{sd_ms}
\end{figure}

The top left panel in figure~\ref{sd_ms} shows the variability in the ensemble mean, measured by the standard deviation over time, versus
lead time. We see that the ensemble mean varies less at longer leads. 
One way this can be understood is that at short lead times the ensemble
is narrow and the ensemble mean can be positioned almost anywhere within the climatological range of the model
forecast. At longer lead times the ensemble is wider, but still has to lie within roughly the same climatological range.
The range of values the mean can take is therefore reduced. 

The upper right panel in figure~\ref{sd_ms} shows the mean of the ensemble spread versus lead time. We see that
the ensemble has low spread at short lead times, and that the spread grows with lead time. The lower left
panel shows the standard deviation of the ensemble spread versus lead time i.e. the variations of the ensemble
spread about the mean. This is smallest at short lead times, shows a maximum, and then reduces at longer
lead times. At very long lead times (longer than these shown here) we would expect it to tend towards
a small positive value, determined by sampling variability. 

Finally, the bottom right panel in figure~\ref{sd_ms} shows the DS ratio versus lead time. Curiously, this ratio
increases almost linearly. At short lead times we see that the ensemble spread is tiny relative to the
variations in the ensemble mean, while at lead times of 8-10 days the standard deviation of the ensemble is
roughly equal to the standard deviation of the ensemble mean.

\subsection{Parameter values and uncertainty}

\begin{figure}[!htb]
  \begin{center}
    \scalebox{0.9}{\includegraphics{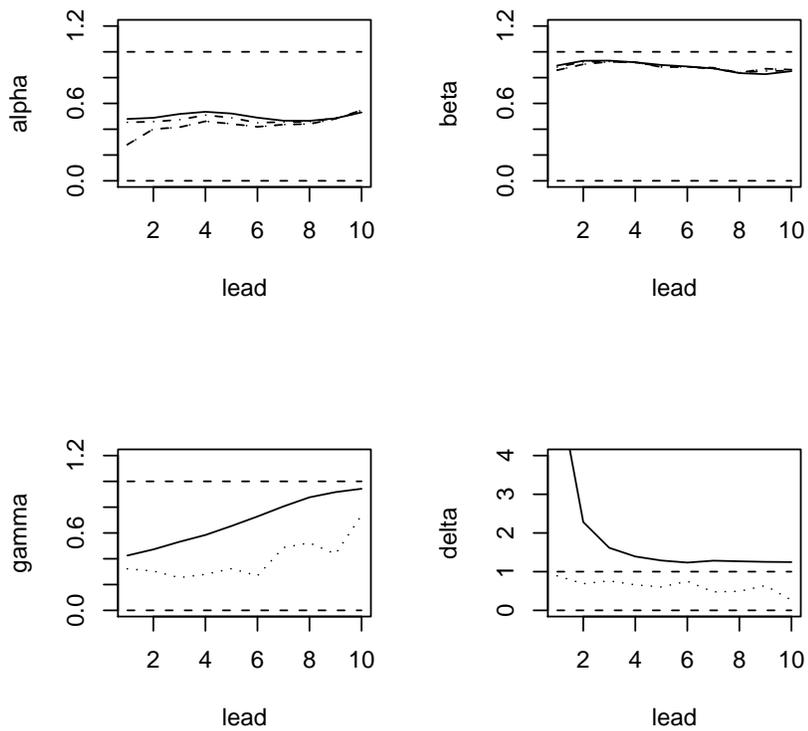}}
  \end{center}
 \caption{Parameter values for the four models discussed in the text. 
 Panels a) and b) show values for the parameters alpha and beta for all four models.
 Panel c) shows values of the parameter gamma for the spread-only and spread-regression models,
 and panel d) shows values of the parameter delta for the spread-scaling and spread regression models.
 In panels a) and b) regression is given by the solid line, spread-only by the dashed line, spread-scaling by
 the dotted line (which cannot be distinguished from the dashed line) and spread-regression by the
 dot-dashed line.
 In panel c) the solid line is regression, and the dotted line is from spread-regression. 
 In panel d) the solid line is from spread-scaling and the dotted line is from spread-regression.} 
 \label{params4}
\end{figure}

We now show the parameters of the four models when fitted to our ensemble data.
The four values of $\hat{\alpha}$ and $\hat{\beta}$ are shown in the upper two panels of figure~\ref{params4}.
The regression and spread-regression models give effectively the same parameter values, while the spread-only
and spread-scaling models give effectively the same values. The latter pair of models give slightly
lower values of $\hat{\alpha}$ than the former especially at short leads.

The two values of $\hat{\gamma}$, for the regression and spread-regression models, are shown in the bottom left panel
of the figure. In the regression model $\hat{\gamma}$ increases roughly linearly with lead time.
The values of $\hat{\gamma}$ from the spread-regression model are somewhat
lower than the values from the regression model, as some of the spread has been taken up by the
$\hat{\delta} s_i$ term. 

The two values of $\hat{\delta}$, for the spread-scaling and spread-regression models, are shown in the bottom right panel
of the figure. In the spread scaling model $\hat{\delta}$ starts off very high and then reduces
to values slightly above one. In the spread-regression model, which does not have to represent the whole spread 
using this term, the values of $\hat{\delta}$ are much lower. 

\begin{figure}[!htb]
  \begin{center}
    \scalebox{0.9}{\includegraphics{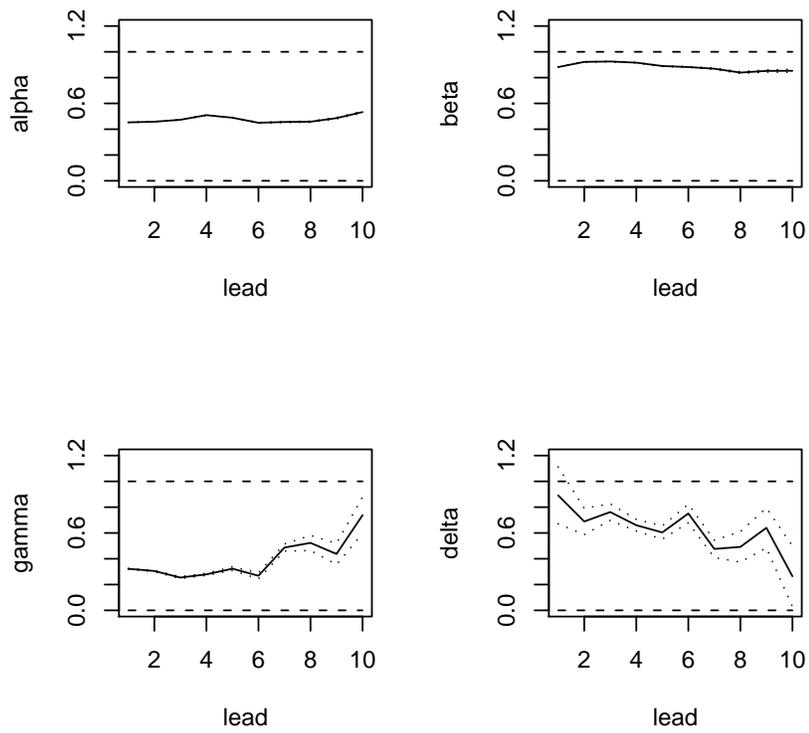}}
  \end{center}
 \caption{Parameter values for the spread-regression model, with confidence limits.} 
 \label{params}
\end{figure}

We now show the four parameters of the spread-regression model but with confidence intervals. These parameter values have already been
shown in~\citet{jewsonbz03a}, but we repeat them here for completeness, and because they tell us a lot about the 
general behaviour of the ensemble in relation to the observations. Figure~\ref{params} shows the four parameters,
confidence intervals representing estimated sampling uncertainty, and the constant levels zero and one.
The confidence intervals are derived in the standard way from the Fisher information.  
For the first two graphs, the confidence levels are so narrow that they are impossible to see. The values
of $\hat{\alpha}$ are significantly different from zero, showing that there is a bias in the forecast, the size of which 
is given by $\hat{\alpha}+\hat{\beta} \overline{m}$. The values of $\hat{\beta}$ show that the uncalibrated ensemble mean varies
in time too much to be an optimum forecast. The ensemble mean needs damping towards zero.
An infinite size perfect ensemble would not need such damping: presumably the reason it is needed 
here is that the ensemble
is not infinite size, and the ensemble members are more correlated with each than they are with reality.

The parameters $\hat{\gamma}$ and $\hat{\delta}$ show greater uncertainty than the parameters $\hat{\alpha}$ and $\hat{\beta}$, presumably
because they are associated with the second moment of the distribution of temperatures rather than the first.
Taking $\hat{\delta}$ first, we see that the variability in the ensemble standard deviation 
(the size of which was shown in figure~\ref{sd_ms}, bottom left panel) 
is too large for the forecast to be optimal beyond lead one.
The factor by which the variability of the spread needs to be reduced lies at around 0.6 up to lead nine. 
The fact that $\hat{\delta}$ is 
significantly different from zero at all lead times is a strong indication that the ensemble spread contains real
information about the uncertainty around the ensemble mean. This need not necessarily be the case 
and raises the hope that the ensemble spread can be used to predict the
uncertainty, and hence improve the probabilistic forecasts that can be made using the ensemble. 
$\hat{\gamma}$ is significantly
different from zero at all leads, and increases with lead. This suggests that, within this model, the ensemble
spread cannot be used on its own to predict forecast uncertainty, and a better prediction (in the likelihood 
maximising sense) would be achieved by using a constant offset too. 

\begin{figure}[!htb]
  \begin{center}
    \scalebox{0.9}{\includegraphics{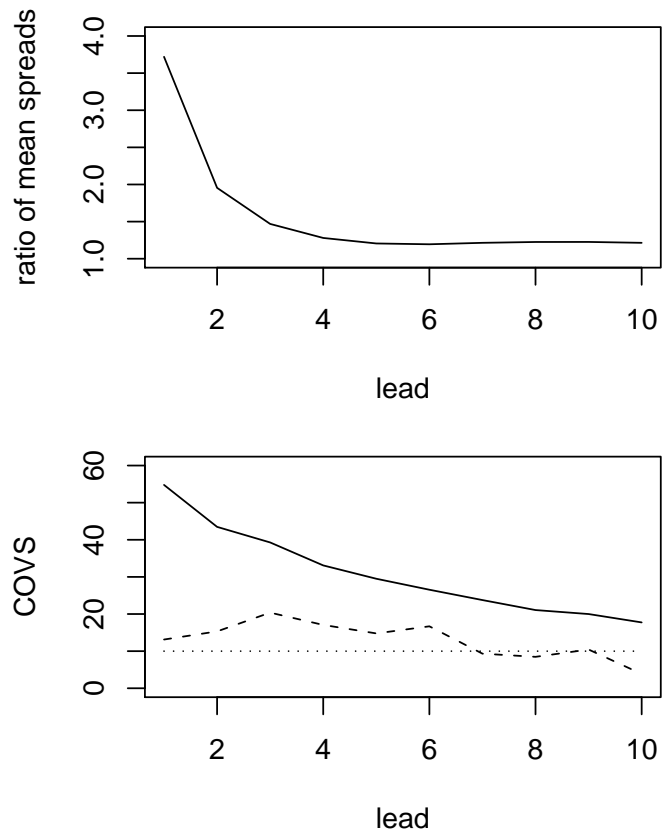}}
  \end{center}
 \caption{Two diagnostic ratios from the ensemble forecasts used in the study.
 The upper panel shows the ratio of the mean of the ensemble spread to the mean
 of the standard deviation in the calibrated ensemble. We see that the uncalibrated
 ensemble spread is too small at all leads, especially leads 1 and 2.
 The lower panel shows the coefficient of variation (COV) of the ensemble spread before (solid line)
 and after (dashed line) calibration. The dotted line is the level of COV that one would expect
 even if the uncertainty was constant, from sampling variability.} 
 \label{ratios}
\end{figure}

We cannot compare the sizes of $\hat{\gamma}$ and
$\hat{\delta}$ in order to get an idea of the relative importance of the terms $\hat{\gamma}$ and $\hat{\delta} s$.
Instead we need to compare the ratios described
in section~\ref{model}. These were also shown in JBZ but are again repeated here because of the important
insight they give to the behaviour of the ensemble. 
The MSR (upper panel of figure~\ref{ratios}) shows that the ensemble spread
is much lower than the actual forecast uncertainty at leads one and two. Combining this with the observation that
the values of $\hat{\delta}$ in the spread-regression model are close to one, 
we see that the variability in the spread is more or less correct, but that a large
constant offset is needed to increase the mean spread up to realistic levels. Were the constant offset term 
$\hat{\gamma}$ not included
in the calibration model then the mean and the variability of the ensemble spread would have to be adjusted
using the same scaling. This would increase the mean spread to realistic levels, but would also increase the variability to
unrealistic levels. This is apparently what happens in the spread-scaling model, and we believe this may also
be what happens in the rank-histogram and best-member calibration methods. 

At longer leads, the average size of the ensemble spread is more realistic, but the values of $\hat{\delta}$ 
from the spread-regression model show us
that the variability of the spread is too large to be optimal. 
The calibration thus replaces much of the ensemble spread with
a constant offset $\hat{\gamma}$ at these leads. 

The second ratio (lower panel of figure~\ref{ratios})
measures the size of the variability of the ensemble spread before and after calibration. 
The horizontal line indicates the value of the COVS that would be expected simply from sampling variability
in an ensemble of this size. 
Before calibration the ensemble spread shows high levels of variability, with a COVS of between 20\% and 50\%,
well above sampling variability.
If this variability were realistic, then one might conclude that predictions of the flow dependent
uncertainty would be very important in making good probabilistic forecasts. However, after calibration the
COVS is much lower, and scarcely above the level of sampling variability. 
We conclude that the optimised uncertainty forecast does not show very much flow dependence. 

The low 
values of the post-calibration COVS are a first indication that the information in the ensemble spread 
is unlikely to improve our probabilistic forecasts a great deal, since the variability in the 
predictable part of the spread is
only a small fraction of the mean spread. 

There are
two possible reasons for the lower values of post-calibration COVS.
First, it could be that the variability in the ensemble spread is very highly correlated with
the variability in the uncertainty, but the amplitude of the variability in the spread
is too large and needs reducing by the factor  $\hat{\delta}$ from the spread regression model to correct
it. 
Secondly, it could be that the amplitude of the variability in the ensemble spread is about right, 
but the correlation between the ensemble
spread and the actual uncertainty is rather low. Reality is probably somewhere between these two
extremes.  
Without a reliable estimator for the linear correlation between the spread and the uncertainty it is rather
hard to distinguish between these two cases. Developing such an estimator is a work in progress. However, given the
ad-hoc methods used to create the ensemble, and the problems with the size of the ensemble
spread at short leads as shown by the MSR, we suspect that the latter case is more likely i.e. that
the variations in the ensemble spread probably have a very low correlation with the actual variations in the 
uncertainty.
This is the optimistic view since if this is true 
then there is a great opportunity for further increases in the skill of ensemble forecasts by increasing our
ability to predict flow-dependent uncertainty, perhaps by improving the methods for generating ensembles.

\begin{figure}[!htb]
  \begin{center}
    \scalebox{0.9}{\includegraphics{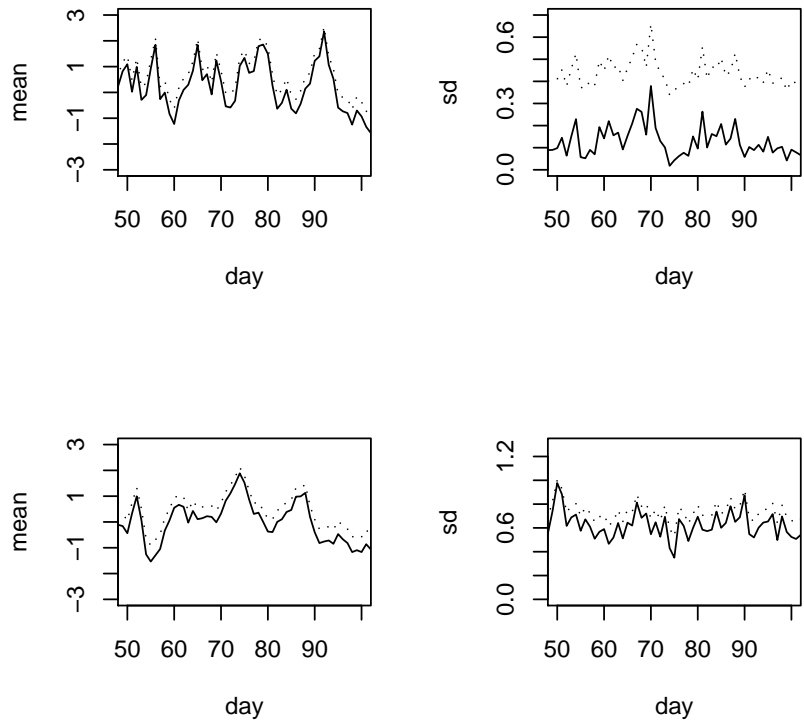}}
  \end{center}
 \caption{Examples of the ensemble mean and standard deviation before (solid line) and
 after (dotted line) calibration. In the left hand panels we see the ensemble mean, and
 in the right hand panels we see the ensemble spread. In the upper panels we see lead
 1 and in the lower panels, lead 5.} 
 \label{calibratedts}
\end{figure}

Finally for this section, we illustrate the effects of the parameters of the spread-regression model
discussed above by showing a sample
of the ensemble mean and spread before and after calibration. Figure~\ref{calibratedts} shows results
for the mean at lead one (upper left), the spread at lead one (upper right), the mean at lead five 
(lower left) and the spread at lead five (lower right). We see that at neither lead does the calibration
of the mean make much difference to the forecast. At lead one, however, the calibration of the spread
is very dramatic, with an increase in the mean level, and a decrease in the variability. At lead five
the effect of the spread calibration is much smaller.

\subsection{Model consistency tests}

\begin{figure}[!htb]
  \begin{center}
    \scalebox{0.9}{\includegraphics{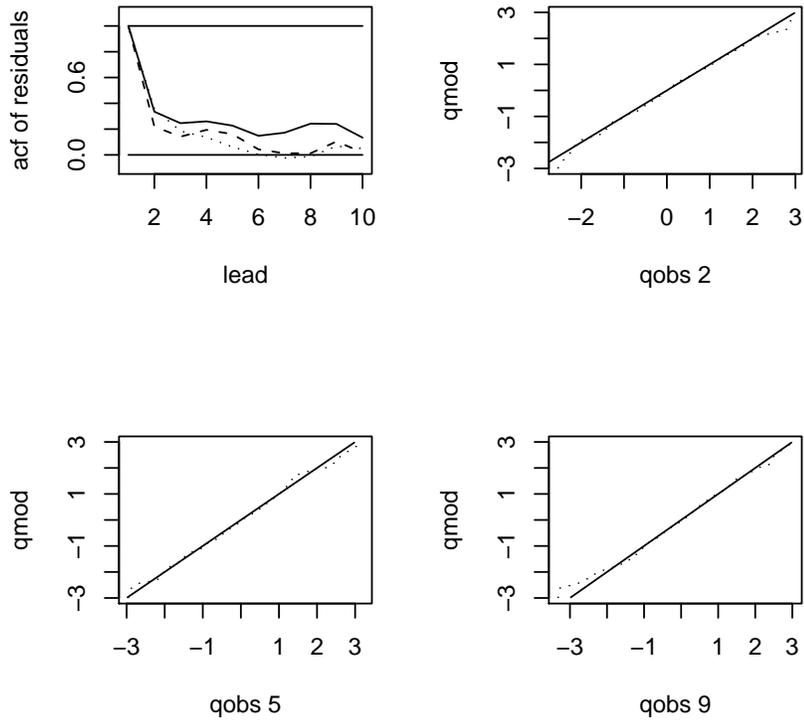}}
  \end{center}
 \caption{Analysis of the residuals from the spread-regression model. 
 The upper left panel shows the ACF of the residuals for 3 different lead times.
 The other panels show QQ plots of the residuals for leads 2, 5 and 9.} 
 \label{resids}
\end{figure}

We now address the question of whether the residuals from the spread-regression model 
are consistent with the 
model assumptions. In particular, we check whether the forecast errors are uncorrelated in time and whether
they are normally distributed. The upper panel of figure~\ref{resids} shows the autocorrelation function
of the forecast errors at three different leads. The results clearly show that the residuals are \emph{not}
uncorrelated in time, but rather show persistent weak positive autocorrelations out to lead ten. This highlights
a definite short-coming of all the models we discuss in this paper, since they have all been fitted under the
assumption that the forecast errors are uncorrelated in time (as have, to our knowledge, all other ensemble
forecast calibration models described in the literature).
Extending the model to overcome this deficiency by modelling $\hat{\Sigma}$ is not entirely
trivial, and is beyond the scope of this paper. It is, however, a high priority for future work.

The lower panel of figure~\ref{resids} shows QQ plots for the residuals versus a normal distribution. It can be
seen that the residuals are in fact very close to normally distributed. In this respect, it would seem that our
models are quite adequate.

\subsection{In sample robustness and skill measures}

We now consider various in-sample tests of robustness and skill. Although out of sample skill measures give us
a better indication of the likely level of skill of future forecasts, in sample skill measures can be very useful
for distinguishing between different calibration models, as we shall see, and for setting an upper limit on the
levels of predictability that we might achieve in real forecasts. In particular, if a complex calibration 
model does not
beat a simple calibration model in in-sample tests, it is unlikely to beat it in out of sample tests, or in a real 
forecasting situation.

\subsubsection{Robustness tests}

\begin{figure}[!htb]
  \begin{center}
    \scalebox{0.9}{\includegraphics{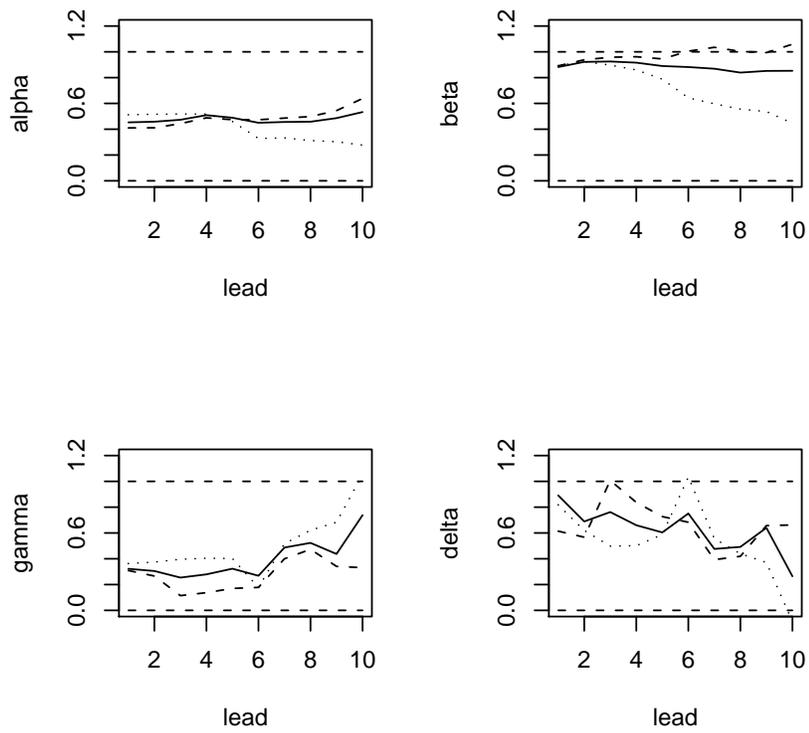}}
  \end{center}
 \caption{A test of the robustness of the parameters of the spread-regression model.
 The solid lines are the all-year parameters, as shown in figure~\ref{params}. The dotted
 lines show the same parameters estimated on the 1st six months of data, and the dashed
 lines show the parameters estimated on the 2nd six months of data.} 
 \label{params12is}
\end{figure}

First, we estimate the parameters of the spread-regression model using the first and second halves of
the data set separately. Figure~\ref{params12is} shows the four parameters estimated using the whole year
of data (solid line), the first half of the data (dashed line) and the second half of the data (dotted line).
The results for the first and second halves of the data are somewhat complementary. For instance, if the whole
data shows a certain bias, and the first half shows a higher bias, then the second half must show a lower bias.
Overall we see reasonable consistency between the parameters estimated on the first and second halves of the
data. For instance, alpha is always greater than zero, beta is less than one up to lead five, gamma is 
always positive, and delta is always less than one but greater than zero. Perhaps surprisingly the differences
between the values for the first and second halves lie well outside the range of sampling uncertainty indicated
by our uncertainty estimates. This is partly because the sampling uncertainty is greater when only using
six months of data, but this cannot explain all the discrepancy. Other possible explanations are that the 
log-likelihood is not close to parabolic at the maximum, or that the system is not non-stationary. Certainly
this last reason is very plausible since we are comparing different times of year, and also because, as discussed
in section~\ref{data}, the forecasting system itself is not held constant in time.

\subsubsection{Skill of the mean}

\begin{figure}[!htb]
  \begin{center}
    \scalebox{0.9}{\includegraphics{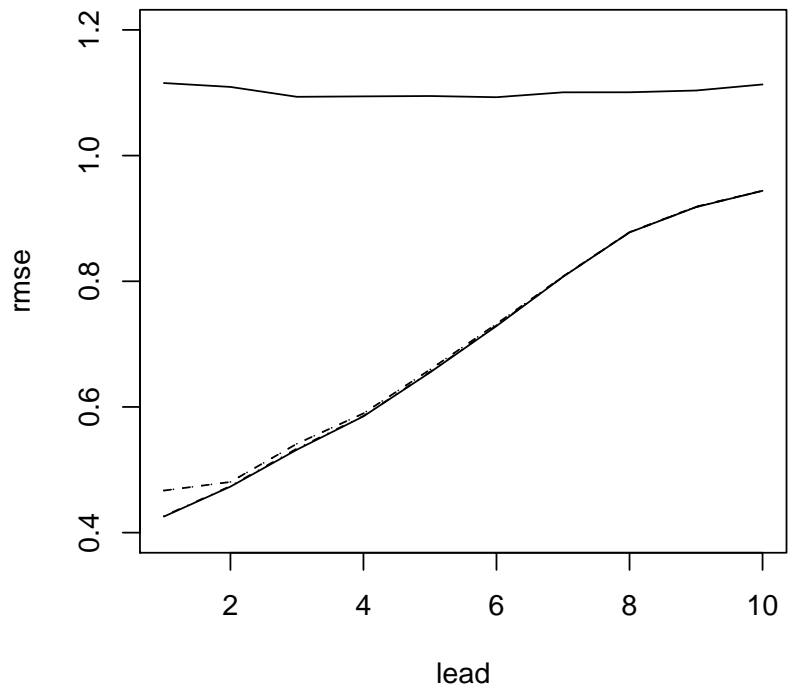}}
  \end{center}
 \caption{
 The RMSEs for the four models described in the text, and climatology.
 Climatology is the roughly horizontal solid line. Regression is the sloping solid
 line. Spread-only is the dashed line, spread-scaling is the dashed line
 (which is indistinguishable from the solid line) and 
 spread-regression is the dot-dashed line (which is also indistinguishable from the solid line).
 } 
 \label{rmse_is}
\end{figure}

We now measure the in-sample skill of our four forecasts.
First, we measure the skill of the mean $\hat{\mu}$ using the
root mean square error (RMSE). Figure~\ref{rmse_is} shows the RMSE values for all four
models and climatology. 
The roughly horizontal line is climatology.
Three of the models (regression, spread-scaling and spread-regression) are almost exactly coincident 
and cannot be distinguished from each other. The fourth model (spread-only) does a bit worse
than the other models.

If all we care about is the skill of the forecast of the
mean temperature, then it does not appear to matter which of regression, spread-scaling or spread-regression we use, 
and since regression is
the simplest as it does need the ensemble spread as an input, then we should use that. However, if we care about
the ability of the forecast to predict the whole distribution, then we need to go beyond the RMSE and look at
whole-distribution measures of skill such as the likelihood in order to distinguish which is the best forecast.

\subsubsection{Probabilistic skill}

\begin{figure}[!htb]
  \begin{center}
    \scalebox{0.9}{\includegraphics{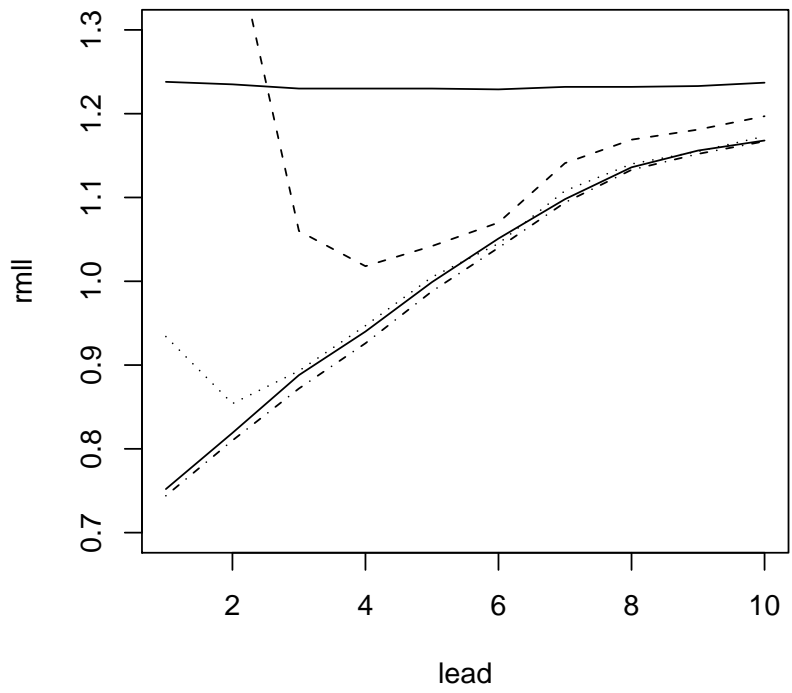}}
  \end{center}
 \caption{
  The RMMLLs for the four models described in the text, and climatology.
 Climatology is the roughly horizontal solid line. Regression is the sloping solid
 line. Spread-only is the dashed line, spread-scaling is the dotted line, and 
 spread-regression is the dot-dashed line.
 }
  \label{rmll_is}
\end{figure}

Figure~\ref{rmll_is} shows the skill of the calibrated ensemble using root mean minus log likelihood
(RMMLL). The solid horizontal line 
shows the RMMLL for climatology. The sloped solid line shows the RMMLL for the regression
model. We see that the regression model gives better probabilistic forecasts than climatology at all lead times.

The dashed line shows the RMMLL for the spread only model.
The results for this model are terrible: at leads one and two the model is even worse than climatology. Since the RMSE
for this model is nearly as good as regression at these leads, we can see that this must be 
almost entirely because the ensemble spread
is an extremely poor prediction for the actual uncertainty. This was perhaps to be expected given that we have already
seen that the ensemble spread at leads one and two is much smaller 
than the uncertainty at those leads (figure~\ref{ratios}, upper panel).
At longer lead times this model improves, but is always distinctly worse than the regression model. We conclude that
the ensemble spread alone is \emph{not at all} a suitable model for the uncertainty in the forecast, and from this point
on will we not consider this model further. 

The dotted line shows results for the spread-scaling model.
This model is always better than climatology, but does a lot worse than regression
at lead one, and slightly worse than regression at all other leads, except perhaps
leads six and nine, where the two models are indistinguishable.
As a result, we will also reject this model from here on since it does worse overall than a simpler model.

Finally the dot-dash line shows the results for the
spread-regression model. This model gives better results than regression at all lead times, and is always the
best of the models. However, it is striking that the results are only very marginally better than the regression
results. One might have hoped that the inclusion of the flow dependent spread via the $\hat{\delta} s$ term would
give a major improvement in the skill of the probabilistic forecast, but this does not appear to be
the case. 

The explanation for this poor relative performance of the spread-regression model 
presumably lies in the small size of the flow dependent changes, as shown in figure~\ref{ratios}, or in
the incorrect model assumption that the residuals are uncorrelated in time. 
It seems unlikely that the second of these is likely to be the main cause of the poor relative performance, 
since this assumption affects
the spread-regression and the regression models equally. We therefore tentatively conclude that the
predictable part of the changes in uncertainty are so small that predicting them
does not make a major difference to the in-sample skill of the probabilistic forecast. 

\subsection{Out of sample tests}

\begin{figure}[!htb]
  \begin{center}
    \scalebox{0.9}{\includegraphics{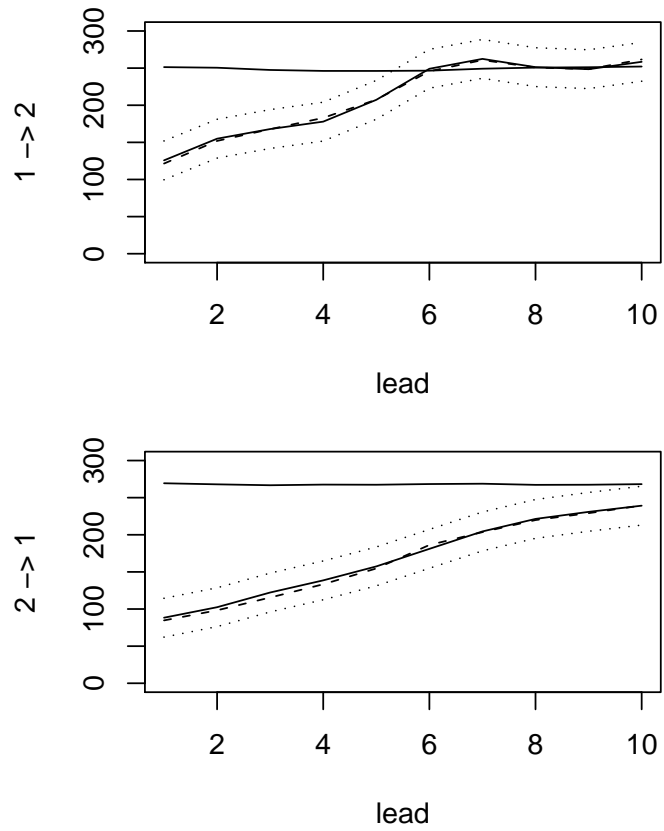}}
  \end{center}
 \caption{The out of sample skill of the spread-regression and regression models, measured
 using RMMLL. 
 The solid line is for regression, the dotted lines are the confidence limits for the regression,
 and the dashed line is for spread-regression.} 
 \label{ll}
\end{figure}

We now move on to out of sample tests. In-sample tests can only give a true indication of the likely level of
predictability if the system is perfectly stationary, and this is never the case in practice because of changes
in the observing system and forecast model, and because of seasonality. 
We have already shown that either the regression model or the spread-regression 
model are the best of the models considered, and that if the spread-regression model is better, it is
only marginally better.
The two particular questions that remain to be answered are a)how well does the regression
model perform versus climatology out of sample and b)does the spread-regression model maintain its (albeit small) 
advantage over the regression model. 
Figure~\ref{ll} shows the RMMLL for calibrated predictions of the second half of the year, where
the calibrating parameters were fitted on the first half of the year (upper panel), and the same for
predictions of the first half of the year made using parameters based on the second half of the year. We feel
that splitting that period into two halves like this gives a fairer indication of the levels of likely
predictability than using leave one out cross-validation (even if whole months are left out) because of the
long-memory characteristic of observed temperatures as documented by~\citet{caballerojb02}. 
We have added confidence intervals onto the 
RMMLL to give some indication of the range of possible values that the RMMLL might take with a different sample.
However, comparison of the two panels, and the fact that the values in the second panel do not lie within
the confidence intervals of the first panel, suggests that these confidence intervals are too narrow. 
As usual, the confidence
intervals are based on the assumption of stationarity of the underlying time series, and this assumption is probably 
wrong. 

The results themselves show that the calibrated probabilistic forecasts contain useful predictability 
up to lead six. The results for the regression and the spread-regression models are virtually 
indistinguishable, and the spread-regression is certainly not significantly better. The tiny advantage 
seen in the in-sample tests is lost when we move to out of sample tests: presumably this is 
due to non-stationarity and sampling error. It would seem that a \emph{much} larger training sample would be needed for
the small signal to be detectable. 

\section{Conclusions}

We have addressed a number of interlinked questions to do with the calibration, assessment and 
predictability of site-specific temperatures using medium range ensemble forecasts. In particular, we have
used a framework which converts an ensemble forecast into probabilistic forecast 
using a simple moment-based calibration model. The calibration model is fitted using
the likelihood, and the ability of the probabilistic forecast to capture the behaviour of observed
temperatures across the whole of the distribution of possible values is also measured using the likelihood. 

Within this framework we have shown that it is possible to detect a correlation between spread and skill,
and that this correlation is robust to splitting the data-set in two. However, the 
size of the predictable variability 
in the uncertainty is small relative to the mean uncertainty.

We have compared a number of specific calibration models within the moment-based calibration framework.
Our results show very clearly that calibrating the ensemble
mean but not calibrating the ensemble spread is a disaster, and at some leads gives a worse probabilistic
forecast than climatology itself. Calibrating the ensemble spread with a single scaling parameter works
better, but does not produce probabilistic forecasts as good as those produced by 
ignoring the ensemble spread completely and using linear regression. 
The only model that compares favourably with linear regression is the spread-regression model in which the 
predicted uncertainty is a linear function of the ensemble spread, including an offset. This model performs
marginally better than linear regression using in-sample tests, but the difference is tiny.
In out of sample tests
the two models perform equally well. Certainly it does not seem to be the case that using the 
extra information available in the ensemble spread improves the out of sample forecasts. This would
imply that, even though the ensemble spread does contain information about the uncertainty, the \emph{amount} of
information is not sufficient to be able to improve a probabilistic forecast based on simple regression.

This conclusion suggests that, given the present forecasts systems, and for this location, 
users of site-specific temperature forecasts who already buy an RMSE-minimising single forecast based on
an ensemble mean
have no reason to buy ensemble forecasts in addition. The value of the ensemble forecast, 
in this case, seems to lie in the mean rather than the spread. 
If they desire probabilistic
forecasts then users can construct them using past forecast error statistics 
and these probabilistic forecasts will be as good as anything based on the ensemble.

One important lesson we can apparently draw from this study is that a significant relationship between 
ensemble spread and skill does not guarantee we can make better probabilistic forecasts by using the spread. 
There is many a slip between cup and lip: 
the spread still has to be calibrated before it can be used 
in a forecast, and performing this calibration is difficult and adds additional uncertainty. 
We have seen that prima-facie reasonable methods
of doing such a calibration, such as the spread-only method, can lead to worse results than simple regression,
which ignores the spread entirely. The only real test of whether spread-skill
relations are of any practical use is to actually produce the final probabilistic forecast with and without
the spread term included, and compare the two in out of sample tests.

There are a number of avenues of future work. Before we finally conclude that the ensemble spread is not a
useful predictor of uncertainty it is important to explore some of the limitations of this study.
One major issue is that we have only used 12 months of ensemble data to train our probabilistic forecasts, 
and only 6 months in our out of sample tests.
Clearly it would be preferable to have much more training data, as long as the data were stationary. 
It is not clear, however, whether, in practice, current forecast systems are held sufficiently constant for
sufficiently long that more data would be better. In fact, it might even be better to use less. 
This depends on 
the size and frequency of changes to the forecasting system, which are unknown (at least to users of the forecasts).
We suspect that the lack of robustness of the parameters in the spread-regression model between the first
and second halves of the year is as much to do with changes in the forecasting system during the second
half of the year as it is to do with seasonality. 
To solve this problem of non-stationarity it is the authors' belief that
ECMWF should run two forecasting systems in parallel, and only change one at a time. Such a system could be used
to ensure that there is always at least 1 year, say, of forecasts from a stationary forecast system. 
Related to this, it would be useful to establish more clearly the relationship between the length of training
data available and the accuracy achieved in fitting the parameters, and whether it is better to fit the 
parameters on annual or seasonal data (this very possibly varies from location to location).

Our analysis made the assumption that the forecast errors are uncorrelated in time, which we
have seen to be untrue. There is a chance that correcting this will improve the skill of the  
probabilistic forecast based on spread-regression to a greater extent than it will improve the regression
based forecast. 
Another assumption was that temperature is normally distributed. Tests of the residuals did not seem to 
contradict this assumption, and so we do not believe that this is a big problem for this particular station.
However, other stations are certainly much less normally distributed than Heathrow: see, for example, the
analysis of Miami temperatures in~\citet{jewsonc03a}. In such cases extending the spread-regression model to
non-normal distributions is a priority: we have already discussed some ways that this might be done in 
section~\ref{extensions}.
It is also of some interest to investigate entirely non-parametric methods, perhaps broadly similar to the
rank-histogram method, that nevertheless do not suffer the spread inflation problem. 
We are in the process of testing such a method.

Our study has focused on temperature: it would be very valuable to attempt a similar analysis for 
precipitation forecasts, in terms of using maximum likelihood methods 
to fit calibration models, using the likelihood to compare models, and assessing the role of spread
in improving forecasts.
It may
well be that ensemble spread is more useful when predicting the distribution of possible precipitation than
it is when predicting the distribution of possible temperatures.

The question of whether major improvements are possible in the prediction of uncertainty remains unanswered.
As discussed above, it may be that the actual uncertainty varies to a much greater degree that our
optimised forecasts of uncertainty, in which case a lot of further improvement would seem to be possible.
We are in the process of developing stochastic analogues of the forecast system which we believe will 
shed light on this question.

Finally, we repeat our main conclusion. We believe in the \emph{prudent use of forecasts}, by which we mean
that users of forecasts should not use a forecast until it has been proven to be better than other simpler
alternatives. We have not been able
to show that using the ensemble spread yields better forecasts than using regression on the ensemble mean alone, and
so we have to advocate that the ensemble spread should not, at this point, be used for making probabilistic
forecasts. However, we would be very happy to see this result refuted, either using the moment-based calibration
framework, or any other.

\section{Acknowledgements}

The author would like to acknowledge Ken Mylne for providing the ECMWF forecasts used in this study, 
Risk Management Solutions for providing us with the observational data, 
Anders Brix for reformatting the forecasts and David Anderson, Anders Brix, Francisco Doblas-Reyes, 
Beth Ebert, Renate Hagedorn 
and Christine Ziehmann for useful discussions on the topic of forecast calibration and validation. 
This research was funded by the author.

\bibliography{jewson}

\begin{thebibliography}{12}
\expandafter\ifx\csname natexlab\endcsname\relax\def\natexlab#1{#1}\fi
\expandafter\ifx\csname url\endcsname\relax
  \def\url#1{{\tt #1}}\fi

\bibitem[Caballero et~al.(2002)Caballero, Jewson, and Brix]{caballerojb02}
R~Caballero, S~Jewson, and A~Brix.
\newblock Long memory in surface air temperature: Detection, modelling and
  application to weather derivative valuation.
\newblock {\em Climate Research}, 21:\penalty0 127--140, 2002.

\bibitem[Casella and Berger(2002)]{casellab02}
G~Casella and R~L Berger.
\newblock {\em Statistical Inference}.
\newblock Duxbury, 2002.

\bibitem[Fisher(1922)]{fisher1922}
R~Fisher.
\newblock On the mathematical foundations of statistics.
\newblock {\em Philosophical Transactions of the Royal Society, A},
  222:\penalty0 309--368, 1922.

\bibitem[Jewson(2003{\natexlab{a}})]{jewson03f}
S~Jewson.
\newblock A note on the use of the word 'likelihood' in statistics and
  meteorology.
\newblock {\em Arxiv}, 2003{\natexlab{a}}.

\bibitem[Jewson(2003{\natexlab{b}})]{jewson03e}
S~Jewson.
\newblock Use of the likelihood for measuring the skill of probabilistic
  forecasts.
\newblock {\em Arxiv}, 2003{\natexlab{b}}.

\bibitem[Jewson et~al.(2003{\natexlab{a}})Jewson, Brix, and
  Ziehmann]{jewsonbz03a}
S~Jewson, A~Brix, and C~Ziehmann.
\newblock A new framework for the assessment and calibration of ensemble
  temperature forecasts.
\newblock {\em ASL}, 2003{\natexlab{a}}.
\newblock Submitted.

\bibitem[Jewson and Caballero(2002)]{jewsonc03a}
S~Jewson and R~Caballero.
\newblock Seasonality in the dynamics of surface air temperature and the
  pricing of weather derivatives.
\newblock {\em Journal of Applied Meteorology}, 2002.
\newblock Submitted.

\bibitem[Jewson et~al.(2003{\natexlab{b}})Jewson, Doblas-Reyes, and
  Hagedorn]{jewsondh03a}
S~Jewson, F~Doblas-Reyes, and R~Hagedorn.
\newblock The assessment and calibration of ensemble seasonal forecasts of
  equatorial pacific ocean temperature and the predictability of uncertainty.
\newblock {\em ASL}, 2003{\natexlab{b}}.
\newblock Submitted.

\bibitem[Jewson and Ziehmann(2003)]{jewsonz03b}
S~Jewson and C~Ziehmann.
\newblock Five guidelines for the evaluation of site-specific medium range
  probabilistic temperature forecasts.
\newblock {\em Arxiv}, 2003.

\bibitem[Molteni et~al.(1996)Molteni, Buizza, Palmer, and
  Petroliagis]{molteniet96}
F~Molteni, R~Buizza, T~Palmer, and T~Petroliagis.
\newblock The {ECMWF} ensemble prediction system: Methodology and validation.
\newblock {\em Q. J. R. Meteorol. Soc.}, 122:\penalty0 73--119, 1996.

\bibitem[Press et~al.(1992)Press, Teukolsky, Vetterling, and Flannery]{NR}
W~Press, S~Teukolsky, W~Vetterling, and B~Flannery.
\newblock {\em Numerical Recipes}.
\newblock Cambridge University Press, 1992.

\bibitem[Roulston and Smith(2003)]{roulstons03}
M~Roulston and L~Smith.
\newblock Combining dynamical and statistical ensembles.
\newblock {\em Tellus A}, 55:\penalty0 16--30, 2003.

\end{thebibliography}

\section{Figures}

\clearpage

\clearpage

\clearpage

\clearpage

\clearpage

\clearpage

\clearpage

\clearpage

\clearpage


\clearpage

\end{document}